# Interaction effects and transport properties of Pt capped Co nanoparticles

A. Ludwig[1,3], L. Agudo[2], G. Eggeler[2], A. Ludwig[3], A. D. Wieck[3], and O. Petracic[1,4]

[1]Institut für Experimentalphysik/Festkörperphysik, Ruhr-Universität Bochum, 44780 Bochum, Germany

[2]Institute for Materials, Ruhr-Universität Bochum, 44780 Bochum, Germany

[3]Lehrstuhl für angewandte Festkörperphysik, Ruhr-Universität Bochum, 44780 Bochum, Germany

[4]Jülich Centre for Neutron Science JCNS-2 and Peter Grünberg Institute PGI-4, Forschungszentrum Jülich GmbH, 52425 Jülich, Germany

**Abstract**

We studied the magnetic and transport properties of Co nanoparticles (NPs) being capped with varying amounts of Pt. Beside field and temperature dependent magnetization measurements we performed $\delta M$ measurements to study the magnetic interactions between the Co NPs. We observe a transition from demagnetizing towards magnetizing interactions between the particles for an increasing amount of Pt capping. Resistivity measurements show a crossover from giant magnetoresistance towards anisotropic magnetoresistance.

**Introduction**

Magnetic NPs attained much interest due to their potential applications e.g. as novel materials with enhanced magnetic, electronic or optical functionalities [1] and for use in magnetic data storage devices [2, 3]. In this context both, an improvement of the thermal stability [4], and a detailed control over the relevant coupling mechanisms is required. Magnetic coupling between NPs can be provided by magnetostatic interactions either of dipolar [5] or multipolar [6] type. Moreover, coupling can also be due to exchange interactions when particles are in direct contact, or due to polarization of a metallic matrix. In the latter case, attempts have been made to study the effect of either embedding or capping (i.e. covering by evaporation-deposition of the capping material on top of the NP) with materials like Al, Cu, Ag, Au, or W [7, 8, 9]. Pt and Pd are of particular interest, because they promote the magnetic polarization by adjacent magnetic moments [10, 11, 12]. In addition, they tend to enhance the effective magnetic anisotropy [13] or induce perpendicular magnetic anisotropy in multilayered systems [14, 15].

Pt is of particular interest because it may help to stabilize the magnetic properties against thermal fluctuations. Here, we present a detailed investigation on the type of magnetic interactions between individual Co NPs in dependence of Pt capping with various nominal thicknesses. Furthermore, the results are complemented by measurements of the magneto-resistance as function of field.

**Sample preparation**

Samples were prepared by sequential ion-beam sputter deposition at room temperature following the sequence [$Al_2O_3$ (3.4 nm)/Co (0.66 nm)/ Pt ($t_{Pt}$)/$Al_2O_3$ (3.4 nm)] at base pressures better than $5 \cdot 10^{-9}$ mbar using highly purified Ar gas. The Co layer with a nominal thickness of 0.66 nm was deposited on an amorphous $Al_2O_3$ buffer layer under a constant deposition angle of 30°. Because of a pronounced non-wetting tendency, Co forms isolated NPs on top of the aluminum oxide layer. This is documented in the scanning transmission electron microscopy (STEM) micrographs shown in Fig. 1. Subsequently, a Pt layer was deposited with different thicknesses $0 \leq t_{Pt} \leq 1.75$ nm for different samples also under this constant deposition angle. The samples were finally covered with an $Al_2O_3$ cap layer under rotation of the substrate to prevent oxidation.

**Structural characterization**

TEM studies were performed to investigate the sample morphology. The images were taken using a FEI Analytical 200 kV FEG-TEM TECNAI F20 S-Twin in the scanning mode. For the TEM investigations the samples were prepared on KBr crystals instead of Si wafers. After the sputter deposition, the KBr crystals were dissolved in water and the film fragments placed on Cu grids.

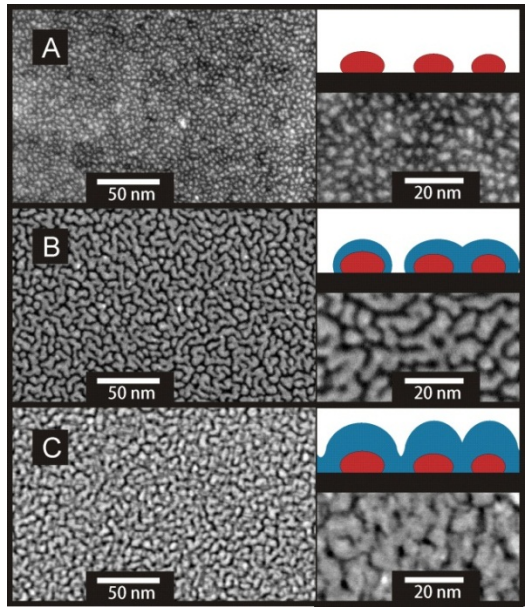

Fig.1: STEM images of uncapped Co NPs (A) and Pt capped systems with $t_{Pt}$ = 0.53 nm (B) and 1.40 nm (C).

The STEM image for the reference sample without Pt (Fig. 1A) reveals isolated Co NPs with a mean diameter of 2.7 nm of more or less spherical shape and a relatively narrow size distribution. One observes for $t_{Pt}$ = 0.53 nm random links between the NPs and for a Pt-capping above 1.40 nm the Pt forms a percolating network.

**Magnetization measurements**

The magnetic properties were studied using a superconducting quantum interference device (SQUID) magnetometer (MPMS, Quantum Design). All measurements were performed with the direction of the applied magnetic field parallel to the sample surface.

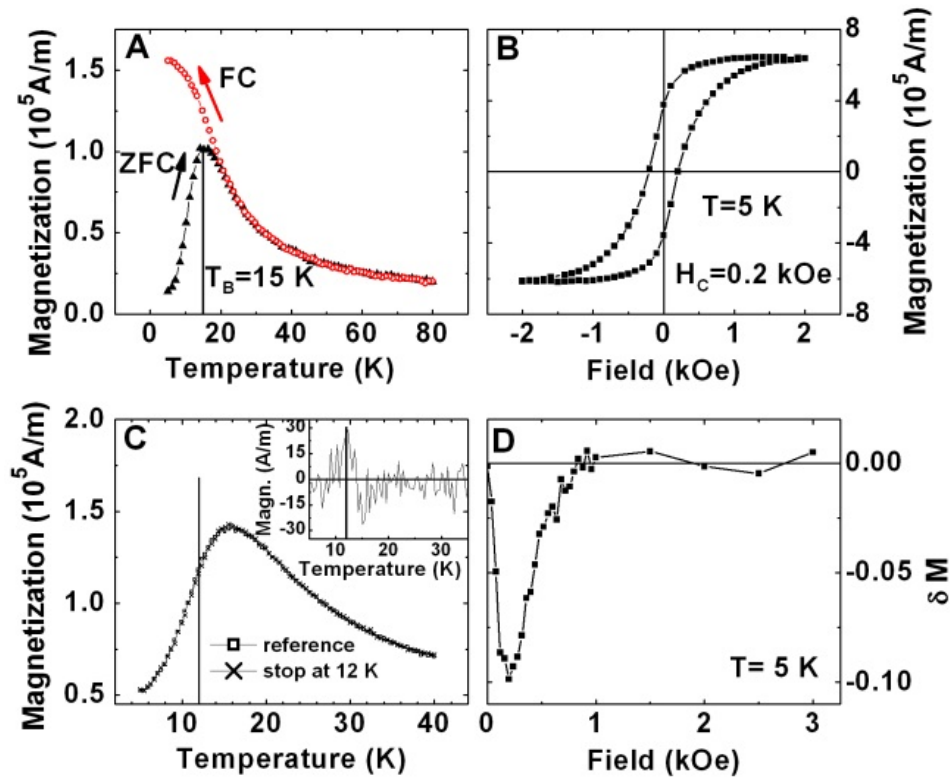

*Fig.2: Magnetization measurements of uncapped NPs: ZFC/FC measurements at H = 20 Oe (A), hysteresis loop at 5 K (B), test for the presence of a memory effect (C) and $\delta M$ curve at 5 K (D).*

Fig. 2A shows the 'zero field cooled' (ZFC) (filled triangles) and the 'field cooled' (FC) (circles) magnetization for the reference sample without Pt capping. After cooling the sample in zero field, the ZFC measurements are performed during heating in a small field of 20 Oe. The shape is typical for superparamagnetic (SPM) particles [16]. For the FC measurement, the magnetization is recorded while cooling the sample in the same field of 20 Oe. The bifurcation temperature, i.e. the temperature where the ZFC and FC curves split, is only slightly above the blocking temperature, supporting the TEM impression of a relatively narrow size distribution

Fig. 2B shows the magnetization hysteresis loop obtained at 5 K. The shape resembles that of an ensemble of Stoner-Wohlfarth particles with an almost random distribution of anisotropy axes. The shape of the magnetization loop hints toward a slightly preferential anisotropy axis along the measurement direction in plane.

Fig. 2C shows a measurement which documents the presence of the memory effect [17, 18]. The memory effect represents direct experimental evidence for spin glass collective inter-particle coupling [18, 19]. To this end, the sample was cooled in zero field to 5 K but with a stop at $T_s$ = 12 K for 20,000 s. Then, the zero field cooling is resumed until 5 K followed by recording the magnetization (ZFC') upon warming in a constant field of 20 Oe. This curve has to be compared to the regular ZFC curve without a stop. In case of the presence of a memory effect, the ZFC' curve will display a dip at $T_s$ recalling the waiting at the stopping temperature. The difference between the two curves (ZFC-ZFC') will then show a peak around $T_s$. The inset of Fig. 2C shows this difference. One indeed observes a peak at the stopping temperature thus indicating superspin glass behavior due to significant dipolar coupling between the NPs [16, 17, 20].

To gain further insight into the type of interaction so-called $\delta M$-curves [21, 22] were measured. The curves are obtained following the expression $\delta M(H) = 2 M_{IRM}(H) - 1 + M_{DCD}(H)$ [21, 22], where $M_{DCD}$ and $M_{IRM}$ are the so-called 'dc demagnetized' and 'isothermal remanent' magnetizations, normalized to the remanent magnetization after positive or negative saturation, respectively. To obtain $M_{DCD}$ the sample is first negatively saturated. Then, a certain positive field, $H$, is applied and the corresponding remanent magnetization value, i.e. at zero field, $M_{DCD}(H)$, is recorded. Following this procedure, the field is then increased successively for each value. The $M_{IRM}(H)$ data is obtained analogously but starting with a demagnetized state. This specific mode has the advantage to minimize the influence of the field step size [23] in contrast to cycling for every field value.

In case of negligible interactions between the particles, $\delta M$ is expected to vanish [24]. It is positive in case of magnetizing and negative in case of demagnetizing interactions. Usually in literature the magnetizing interactions are attributed to exchange interactions and the demagnetizing interactions to dipolar coupling [21]. To reduce the effect of superspin fluctuations, the measurements were performed at $T$ = 5 K.

The $\delta M$-curve of the sample without Pt capping (Fig. 2D) encloses a negative area which implies a *demagnetizing* coupling due to frustrated dipolar interactions. This supports the results obtained by the memory measurements.

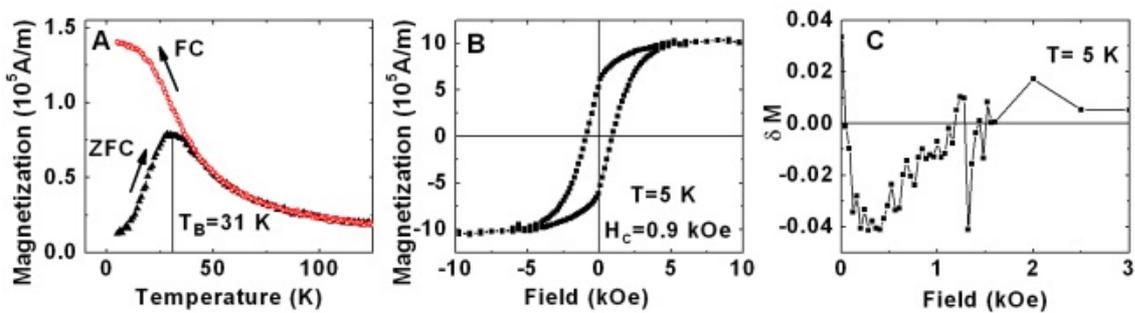

*Fig.3: Magnetization data for the capped system with $t_{Pt}$ = 0.18 nm: ZFC/FC measurements at H = 20 Oe (A), hysteresis loop at T = 5 K(B) and $\delta M$ curve at T = 5 K (C).*

For a Pt capping of 0.18 nm, the $\delta M$ curve area is still negative, however, the maximum amplitude is reduced by more than a factor of 2 (Fig. 3C). Furthermore, the peak is broadened towards larger field values which reflects the increased coercive field as obtained by the hysteresis loop (Fig. 3B). The ZFC/FC measurement (Fig. 3A) shows an increased blocking temperature, which is caused by the increase in the effective magnetic volume and therefore an increased energy barrier of the NPs [25].

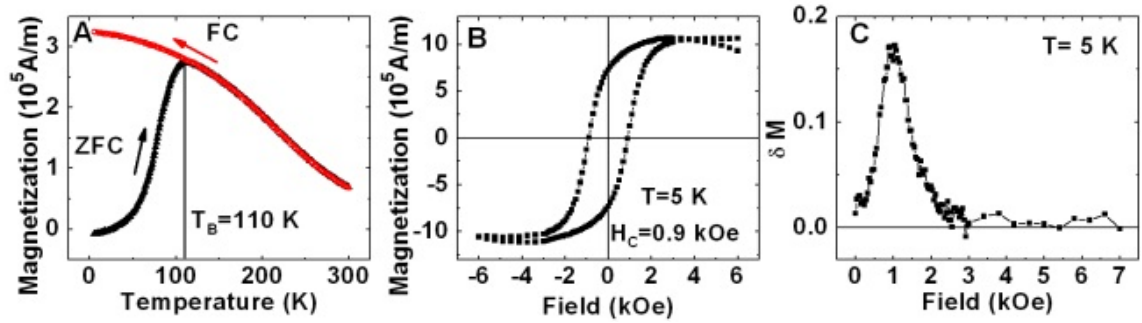

*Fig.4: Magnetization measurements for $t_{Pt}$ = 0.53 nm: ZFC/FC measurements at H = 20 Oe (A), hysteresis loop at T = 5 K (B) and $\delta M$ curve at T = 5 K (C).*

For the Co NPs capped with 0.53 nm Pt, an increase in the blocking temperature of up to $T$ = 110 K is observed (Fig. 4A). Although the shape of the ZFC and FC measurements is still similar to those of the previous systems the amplitude is more than two times larger. This finding indicates a more parallel alignment of the superspins. This is consistent with the $\delta M$ curve (Fig. 4C) displaying a positive area and consequently suggests that there is a *magnetizing* interaction between the NPs. As seen in the STEM images (Fig. 1), the Co NPs capped with 0.53 nm Pt are no longer separated but form clusters of several NPs interconnected by Pt bridges. The crossover in the nature of interaction is therefore most likely mediated by this type of bridging.

Note that from a comparison of the amplitudes of the $M(T)$ curves on the samples with $t_{Pt} \leq 0.53$ nm we infer that no significant alloying of Co and Pt occurs in our system [25]. I.e., the amplitude of the ZFC-curves recorded at the same field value does not show the systematic decrease with increasing Pt capping. In fact, even an increase can be observed in our case [25]. This is in contrast to similar studies in literature, where alloying was found [7]. The alloying was signified by a systematic reduction of the magnetization with increasing capping.

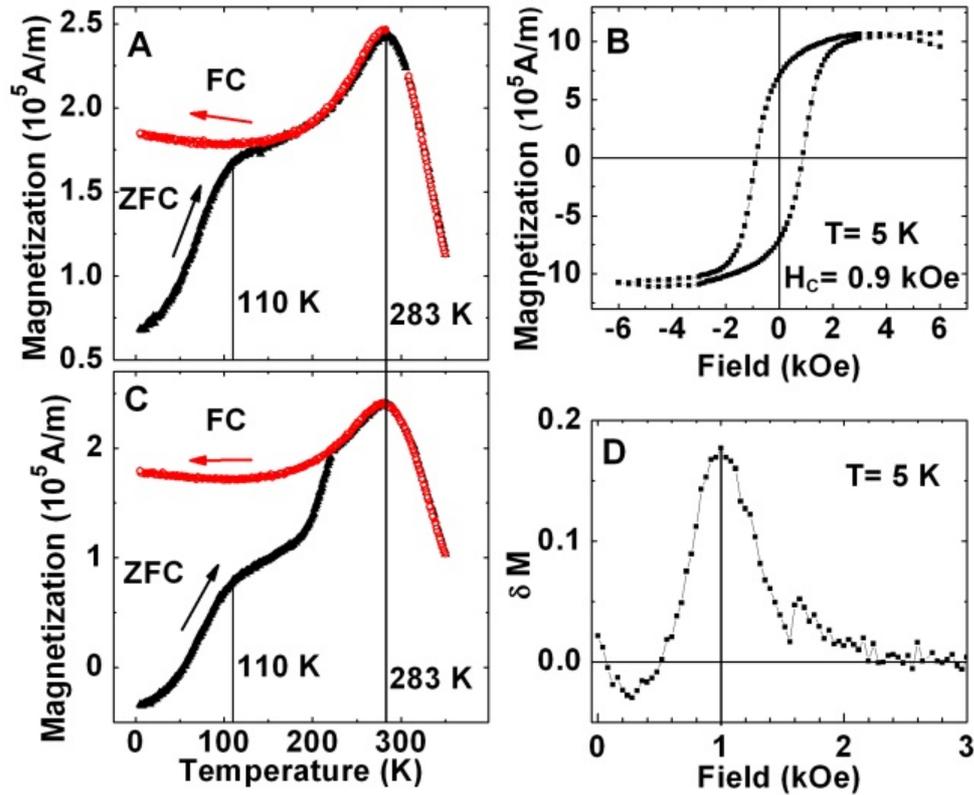

*Fig.5: Magnetization measurements for $t_{Pt}$ = 0.70 nm: ZFC/FC curves at H = 20 Oe with a field of H = +1 kOe (A) and H = -1 kOe (C) applied before the initial cooling, hysteresis loop at T = 5 K (B) and $\delta M$ curve at T = 5 K (D).*

A further increase in the amount of Pt leads to a qualitatively different behavior. The *M(T)* measurements are dependent of the sample history and a remanent magnetization is observed even for temperatures up to 350 K after letting the system relax for 30 min. Hence, we modified the procedure of the ZFC/FC measurements: *Before* zero-field cooling, a field of +1 kOe or -1 kOe was applied at 350 K to saturate the sample positively or negatively, respectively. Then the field was removed and the sample relaxed for 30 min at 350 K. By applying the field, all NP moments align in the direction of the field. Assuming a superposition of strongly and slightly coupled NPs, the relaxation time may not influence the orientation of the strongly coupled NP moments, but results in the fact that only the slightly coupled particles change their moments and switch to energetically more favorable directions.

Then the sample was cooled to $T$ = 5 K in zero magnetic field and subsequently the ZFC/FC curves were recorded as before. In Fig. 5A and C the results for the Co-NPs capped with 0.70 nm are shown for both the positively and negatively saturating field applied at 350 K. Both ZFC-curves show a bump around 110 K which corresponds to the blocking temperature of the NPs capped with 0.53 nm Pt. The amplitude of this contribution is nearly the same in both measurements, suggesting that this contribution is independent of the sample history and therefore based on the switching of *independent* clusters of NPs or individual NPs similar to the system with $t_{Pt}$ = 0.53 nm.

Further similarities between the two measurements shown in Fig. 5A and C are the increase in the magnetization up to a peak at $T$ = 283 K, followed by a relatively strong decrease for increasing temperatures and a decreasing FC curve for decreasing temperatures. The peak might indicate a completely different type of coupled behavior, e.g. of antiferromagnetic type. However, the sharp increase in the magnetization, present only in the negatively saturated protocol, might indicate domain wall movement type of switching in ferromagnetically correlated subsystems. Completely ferromagnetic coupled behavior is observed for 1.58 nm ≥ $t_{Pt}$ ≥ 1.05 nm [25]. Hence, the samples with 0.70 nm Pt (and also 0.88 nm) show properties which can be attributed to a transition regime with mixed yet elusive behavior.

The hysteresis loop (Fig. 5B) shows basically Stoner-Wohlfarth type of behavior with a rounded S-shaped loop. The $\delta M$ curve is also similar to the one for 0.53 nm Pt, viz. showing an overall positive area.

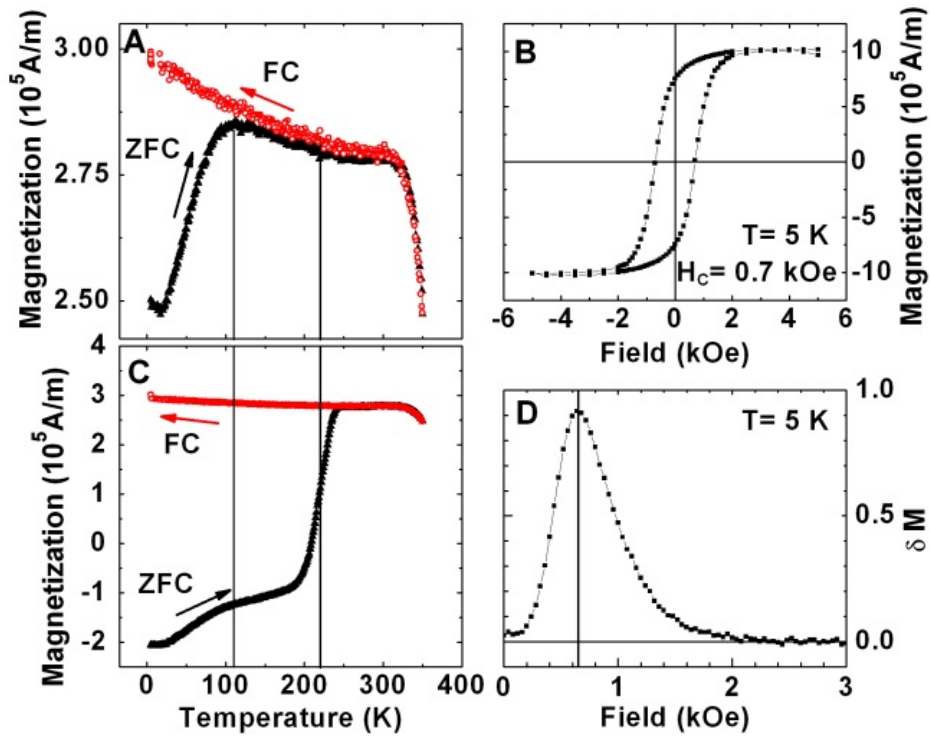

*Fig.6: Magnetization measurements for $t_{Pt}$ = 0.88 nm: ZFC/FC curves at H = 20 Oe with a field of + 1 kOe (A) and -1 kOe (C) applied before the initial cooling, hysteresis loop at 5 K (B) and $\delta M$ curve at 5 K (D).*

For the sample with 0.88 nm Pt capping layer, the ZFC/FC curves, corresponding to the two different field treatment protocols at 350 K, (Fig. 6 A, C) show similarities but also significant differences to the case before. Again, a bump at 110 K is observed indicating the response of uncoupled clusters of particles or single particles as discussed above. This contribution is smaller compared to the rest of the curve which is consistent with the assumption that an increased Pt capping couples an increased number of NPs. Also, the relatively steep increase at ~210 K in the ZFC curve in Fig. 6C resembles the one in Fig. 5C, however, with a larger jump. We assume switching inside ferromagnetically correlated regions via domain wall motion. This behavior is similar to the completely ferromagnetic case shown below, however, limited only to finite regions inside the sample. Also, a relatively sharp decrease of both ZFC and FC curves is observed for increasing temperatures. Compared to the previous case, this occurs at a higher temperature, i.e. ~320 K. In the FC curve of Fig. 6A also, a small peak is found at ~310 K.

In contrast to the previous case, both ZFC and FC curves do not show a decrease with decreasing temperatures. Here again, we assume an intermediate regime between superparamagnetic-like and completely ferromagnetic-like behavior, with a mixed response due to partially ferromagnetically coupled regions, still relatively strong dipolar interactions and antiferromagnetic correlations.

The peak in $\delta M$ vs. $H$ (Fig. 6D) shifts to smaller field values compared to the previous case with a maximum at $H$ = 0.65 kOe. Additionally, the amplitude is larger with a maximum value of 0.9, which indicates stronger magnetizing (i.e. ferromagnetic) coupling, corroborating the assumption of larger FM-coupled regions. The FM coupling is mediated by Pt bridges connecting the NPs. The underlying mechanism is most likely polarization of the Pt by the Co and is possibly also influenced by RKKY-interactions [26, 27]. The hysteresis loop (Fig. 6B) shows an increased square-like shape, which also confirms the presence of FM-regions switching by domain wall motion.

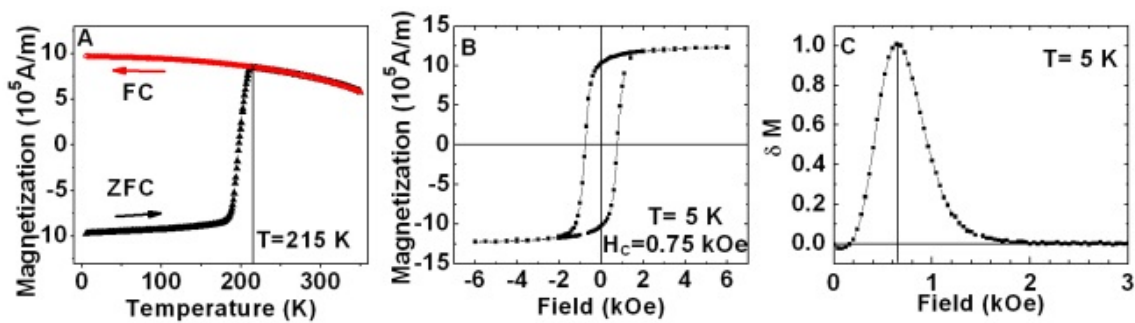

*Fig.7: Magnetization measurements for $t_{Pt}$ = 1.05 nm: ZFC/FC curves at H = 20 Oe with a field of -1 kOe applied before the initial cooling(A), hysteresis loop at 5 K (B) and $\delta M$ curve at 5 K (C).*

For even larger amounts of Pt deposited on the Co NPs (1.05 nm ≤ $t_{Pt}$ ≤ 1.58 nm), the ZFC/FC- curves reveal a completely different scenario. In Fig. 7A the results for a Pt-capping of 1.05 nm are shown for the case of a negative field applied before cooling. (For the case of applying a positive field the ZFC/FC curves collapse onto the FC curve.) One finds that the amplitude of the ZFC-curve is identical to the FC-curve, however, with opposite sign. Both curves basically represent the order parameter of a FM. Since the FC curve is obtained at relatively small fields, it represents the order parameter curve approximately. The ZFC curve also represents the order parameter curve but with intermediate switching from negative to positive amplitude. This occurs exactly at that temperature, where the applied field ($H$= 20 Oe) matches the coercive field of the system.

The FC-curve can be modeled by a semi-empirical formula developed for ferromagnets [25, 28]. The fit reveals a critical temperature of 409 K [25]. Hence, within this range of Pt-capping, the NPs are strongly ferromagnetically coupled. This is also confirmed by the relatively square-like hysteresis loop (Fig. 7B) and the positive $\delta M$ curve shown in Fig. 7C.

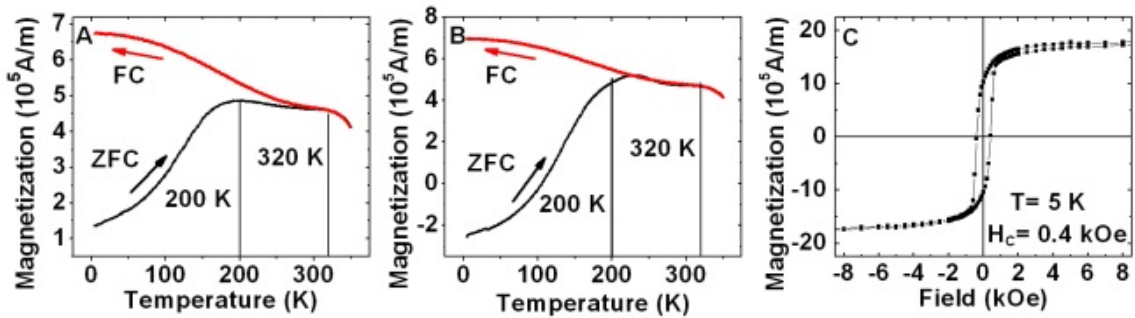

*Fig.8: Magnetization measurements for $t_{Pt}$ = 1.75 nm: ZFC/FC curves at H = 20 Oe with a field of + 1 kOe (A) and -1 kOe (B) applied before the initial cooling, hysteresis loop at 5 K (C).*

Fig. 8 shows the results obtained for the system with the largest Pt-capping studied here, i.e. $t_{Pt}$ = 1.75 nm. Although the FM-like coupling is observed nearly unchanged over a wide range of Pt capping ($t_{Pt}$= 1.05 - 1.58 nm), the distinctive switching behavior is absent for $t_{Pt}$ = 1.75 nm (Fig. 8A, B). Instead, the ZFC/FC curves have a superparamagnetic-like shape for temperatures below 250 K. However, a difference is found in the amplitude of the ZFC curves at $T$ = 5 K, which suggests an interaction promoting parallel alignment of the NP-superspins. The additional peak at $T$ = 320 K might indicate a residual ferromagnetically-coupled behavior of the entire system. However, the excess amount of Pt obviously cannot be polarized completely as in the previous case. We infer that two subsystems emerge, i.e. (I) Co NPs strongly coupled to the surrounding Pt, thereby weakly coupling the NPs and (II) 'excess' Pt being dragged by the first subsystem, however, being not completely polarized. The $M(H)$ hysteresis loop (Fig. 8C ) shows a reduced coercivity and a more gradual switching. The rounded shape might result from the paramagnetic-type of signal of the subsystem II.

**Resistivity measurements**

To study the impact of the magnetic properties on the electronic transport behavior, we fabricated Hall bar structures, using Au contacts. The direction of the applied field lies within the sample plane for all measurements. The direction of the current was longitudinal as well as transversal to the applied magnetic field.

Fig. 9 A, B shows the magnetoresistance (MR) measurements for the sample capped with $t_{Pt}$ = 1.40 nm. For the transversal configuration, the resistance decreases with increasing field values, whereby the maxima are shifted to 0.3 kOe significantly below the value of the coercive field of 0.5 kOe as obtained from magnetometry. In the longitudinal case, the resistance increases for larger magnetic field values with a maximum amplitude of ~0.3%, which is nearly the same as for the current perpendicular to the magnetic field. This finding for both geometries can consistently be ascribed to the regular anisotropic magnetoresistance (AMR) effect. This then implies that the entire system (Co NPs and Pt network) behaves as one FM system. (Note that this sample has been measured in a closed-cycle cryostat at 5 K, whereas the other two samples were measured in a helium-bath cryostat at 4.2 K.)

For $t_{Pt}$= 0.88 nm (Fig. 9 C, D) the shapes of the MR-curves are only slightly different. The maximum amplitude is no longer the same for both configurations and is reduced to 0.13% for longitudinal and 0.06% for transversal configuration. The field values of the maxima correspond to the coercivity of the magnetization hysteresis. Also, in this case, the transport behavior is dominated by the AMR effect. Consequently, the entire system is again electrically percolated and behaves as one FM as in the case before.

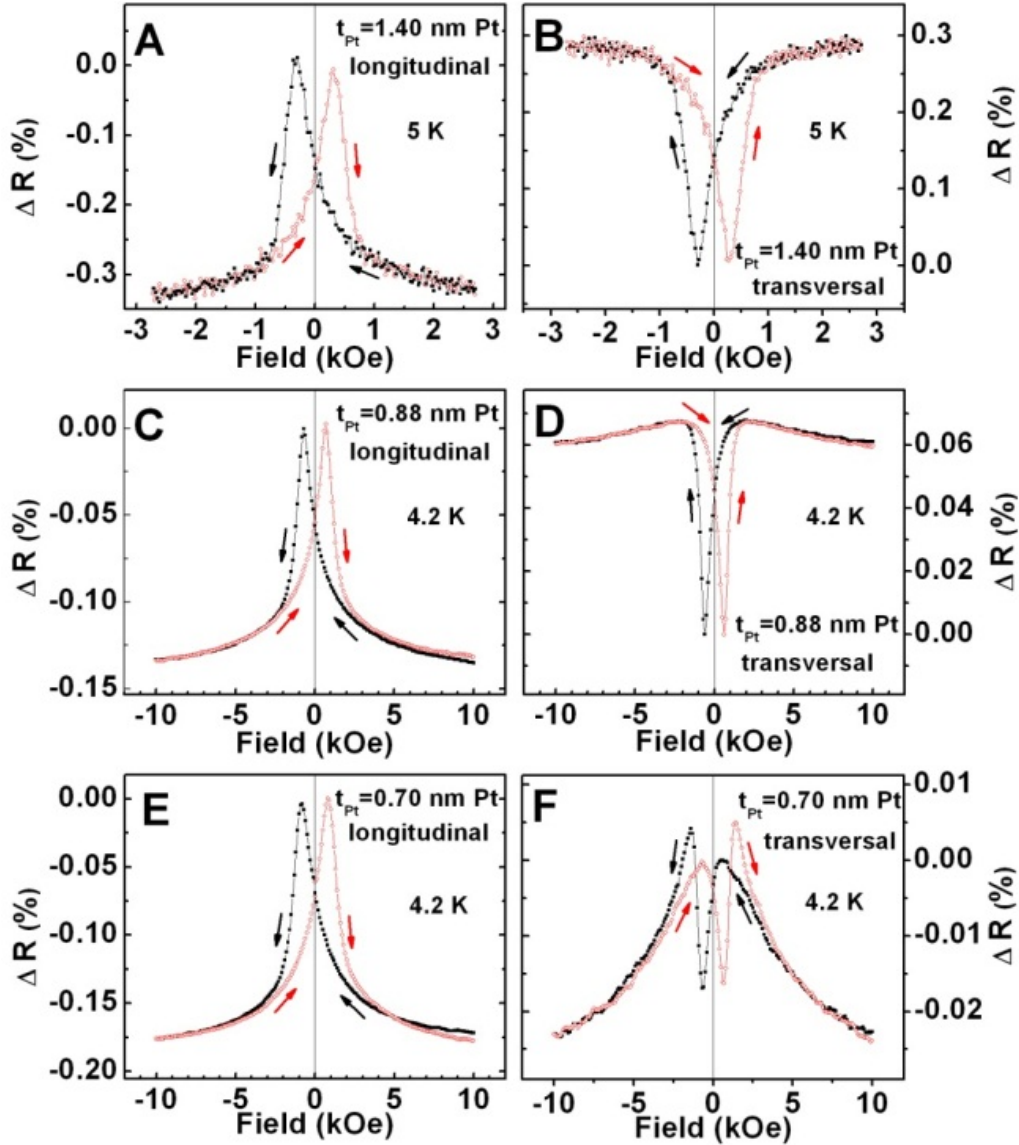

*Fig. 9: Relative Magnetoresistance curves as function of applied field with current direction parallel to field (A,C,E) and transversal to field (B,D,F) for $t_{Pt}$ = 1.40 nm (A,B), 0.88 nm (C,D) and 0.70 nm (E,F) measured at T = 4.2 K and T = 5 K, respectively. The sheet resistance varies from 0.9 kΩ over 0.6 kΩ to 0.1 kΩ for increasing Pt coverage.*

For the system with $t_{Pt}$=0.70 nm (Fig. 9 E, F), the measurements performed with a current parallel to the magnetic field are very similar to those performed for the sample capped with 0.88 nm Pt. The maximum amplitudes are slightly shifted towards higher field values, which is consistent with the coercive fields obtained from magnetization measurements. The curve for the transversal geometry, however, shows an additional dip. This might be due to the influence of a tunneling conductance due to partially not electrically connected NPs via the Pt. As seen in the TEM image (Fig.1) for $t_{Pt}$=0.53 nm the NPs are connected via the Pt and form small chains or clusters.

Systems with even less Pt show resistances of the order of MΩ-GΩ and are thus non-percolated. This type of granular systems, with tunneling type of conductance have been investigated extensively in the literature before [29, 30] and thus were not in the focus of the present investigation.

**Conclusion**

We have investigated the influence of Pt capping on the magnetic and transport properties of Cobalt nanoparticle ensembles. With increasing Pt capping we find a crossover from demagnetizing interactions (due to dipolar inter-particle coupling) to magnetizing interactions (mediated via a percolating network of Pt-bridges). This is evidenced using in particular $\delta$M-curves. The systems with relatively small amounts of Pt ($t_{Pt}$ ≤ 0.53 nm) behave superparamagnetic-like. For capping thickness values 0.7 nm ≤ $t_{Pt}$ ≤ 0.88 nm, an intermediate range is observed, i.e. a superposition of superparamagnetic and antiferro- and ferromagnetic correlations exists. For relatively large capping (1.05 nm ≤ $t_{Pt}$ ≤ 1.58 nm), completely ferromagnetic coupling between the Co NPs due to a percolated Pt network is found. For the largest investigated capping with $t_{Pt}$ = 1.75 nm, an excess of Pt results in a decoupling of strongly correlated Co-Pt regions and less coupled Pt regions coexist. Most of the magnetic signatures are reproduced in the electronic transport measurements.

**Acknowledgment**

We thank Hartmut Zabel for valuable discussions and support. Moreover, financial support by the Materials Research Department of the Ruhr-University Bochum is acknowledged.